\documentclass{llncs}

\usepackage{hyperref}

\begin{document}

\title{Multi-Agent Programming Contest 2010 \\ --- \\ The Jason-DTU Team}

\author{J{\o}rgen Villadsen \and Niklas Skamriis Boss \and Andreas Schmidt Jensen \and \\ Steen Vester}

\institute{Department of Informatics and Mathematical Modelling \\
Technical University of Denmark \\
Richard Petersens Plads, Building 321, DK-2800 Kongens Lyngby, Denmark}

\thispagestyle{plain} \maketitle \pagestyle{plain}

\bigskip

\begin{abstract}
We provide a brief description of the Jason-DTU system, including the methodology, the tools and the team strategy that we plan to use in the agent contest.

\

Updated 1 October 2010: Appendix with comments on the contest added.
\end{abstract}

\bigskip

\section{Introduction}

\begin{enumerate}
\item
The name of our team is Jason-DTU. We participated in the contest for the first time in 2009 where we finished number 4 out of 8 teams \cite{Boss+2010}.
\medskip
\item
The members of the team are as follows:
\begin{itemize}
\smallskip
\item J{\o}rgen Villadsen, PhD
\smallskip
\item Niklas Skamriis Boss, MSc
\smallskip
\item Andreas Schmidt Jensen, MSc
\smallskip
\item Steen Vester, BSc (currently MSc student, new in the team this year)
\smallskip
\end{itemize}
We are affiliated with DTU Informatics (short for Department of Informatics and Mathematical Modelling, Technical University of Denmark, and located in the greater Copenhagen area).
\medskip
\item
We use the Jason platform, which is an interpreter for AgentSpeak, an agent-oriented programming language \cite{Bordini+2007}.
\medskip
\item
The main contact is associate professor J{\o}rgen Villadsen, DTU Informatics, email: \email{jv@imm.dtu.dk}
\medskip
\item
 We expect that we will have invested approximately 100 man hours when the tournament starts.
\end{enumerate}

\section{System Analysis and Design}

\begin{enumerate}
\item
We intend to use three types of agents: a leader, a scout and the regular herders.
The leader is a herder with extra responsibilities and the scout will initially explore the environment.
We do not use a specific requirement analysis approach.
\medskip
\item
We design our system using the Prometheus methodology as a guideline \cite{Padgham+2007}.
By this we mean that we have adapted relevant concepts from the methodology, while not following it too strictly.
\medskip
\item
The agents navigate using the A* algorithm \cite{Russell+2003}.
We also implement algorithms that enable the agents to move in a formation and to detect groups of cows.
\medskip
\item
Communication is primarily between individual agents and the leader.
Each agent has a role based on their type.
Coordination is done by the leader.
\medskip
\item
We have chosen to have a centralized coordination mechanism in form of a leader.
\end{enumerate}

\section{Software Architecture}

\begin{enumerate}
\item
We use the Jason platform and the AgentSpeak programming language to specify the goals an agent must pursue.
Furthermore, we are able to use Java using so-called internal actions.
\medskip
\item
We use the architecture customization available in the Jason platform.
Each agent is associated with an agent architecture which contains basic functionality such as connecting to the server and sharing knowledge.
This enables us to implement and custumize the agents in a rather elegant way.
\medskip
\item
We use the Jason platform within the Eclipse IDE.
\end{enumerate}

\section{Agent Team Strategy}

\begin{enumerate}
\item
We use mainly the A* algorithm to avoid obstacles and we do not use any algorithms for opponent blocking at the moment.
\medskip
\item
The team leader handles coordination.
Each herder will get delegated a position from which it must herd.
\medskip
\item
We do not employ a distributed optimization technique, however, the leader chooses an agent which is currently closest to the goal.
\medskip
\item
All knowledge is shared between the agents.
This means that every agent knows everything about the environment.
Furthermore, each agent communicates with the leader.
\medskip
\item
We plan to consider a more autonomous and decentralized approach where each agent is able to decide without having to ask the leader.
\medskip
\item
Our agents do not perform any background processing while the team is idle,
i.e. between sending an action message to the simulation server and receiving a perception message for the subsequent simulation step.
\medskip
\item
We do not have a crash recovery measure.
\par
\smallskip
Whereas classical multi-agent systems have the agent in center, there have recently been a development towards focusing more on the organization of the system.
If time permits we would like to investigate the pros and cons of a more organizational approach \cite{Jensen2010}.
\end{enumerate}

\

\section*{Acknowledgements}

Thanks to Mikko Berggren Ettienne (BEng student) for joining the team.

\vfill

\begin{center}
More information about the Jason-DTU team is available here:
\\[3ex]
\url{http://www.imm.dtu.dk/~jv/MAS}
\end{center}

\newpage

\section*{Appendix}

We gained the insight about the practical use of multi-agent systems
that domain specific knowledge is quite important in a multi-agent
system like the one in the contest. General concepts of search
algorithms, belief sharing, communication and organization are
important too and provide a solid basis for a good solution. However,
we think that domain specific topics such as understanding cow
movement, refinement of herding strategy, obstruction of enemy goals
etc. were even more important to obtain success. We definitely spent
most of our time doing domain specific refinements and performance
tests.

The scenario had some nice properties like uncertainty about the
environment, nondeterministic cow movement and the need for agent
cooperation to obtain good herding results. These properties made sure
that good solutions were non-trivial and gave motivation for
experimenting with a lot of different approaches. Interaction with an
enemy team is also very interesting. Though, we feel that care should
be taken when designing a scenario so it will not be too easy to
implement a near-perfect destructive strategy which will ruin the
motivation for pursuing other ideas.

We used a centralized structure with one leader delegating targets to
all other agents which gave an overall control of our team. The leader
divided the agents into groups which had different purposes. For
example we had a couple of herding groups and a group responsible for
making life harder for our opponents' herders. Originally we used
fixed groups (with fixed sizes) of agents with quite static
responsibilities. We learned that it can be important for agents to
switch roles if the environment acts in a way that makes this
preferable. We did some experiments with this when forming groups of
herders. In some cases the environment (and our agents) acted in such
a way that it was more optimal for agents to switch to other groups
than to stick with the predefined groups.

We have a few ideas for potential extensions of the
cow-and-cowboys-scenario. One issue is that changes should be made so
that a destructive approach will not be as beneficial as it was this
year. One suggestion is to restrict the number of agents that can be
in the corral of the enemy at any time, for example by automatically
teleport additional agents to their own corral, but of course this
makes the scenario quite unrealistic. Some other ideas are to let the
cows be controlled by one or more teams and perhaps allowing the
number of cows to increase or decrease over time.

We prefer to stay with a variant of the current scenario for the
coming year but eventually a less toy-like scenario might be
introduced. Perhaps some kind of scenario within health, food, energy,
climate or engineering would be possible. Alternatively one could move
towards computer games (say, World of Warcraft).

We think that the contest was organized very well and that the
information regarding protocols and rules were quite clear. Even
though several members of the team had not participated in an event
like this before we did not experience any problems communicating with
the servers and we feel that the information level overall was quite
good. The live chat was also a positive feature.

\end{document}